%% file: qlg_communication.tex
\documentclass[final,5p,times,twocolumn]{elsarticle}

\usepackage{amsmath}
\usepackage{booktabs}
\usepackage{array}
\usepackage{threeparttable}
\usepackage{xcolor}
\graphicspath{{figs/}}
\usepackage{orcidlink}
\usepackage{hyperref}
\hypersetup{
    colorlinks=true,
    linkcolor=blue,
    filecolor=magenta,
    urlcolor=cyan,
    citecolor=blue,
    breaklinks=true
}

\newcommand{\K}{K}
\newcommand{\hw}{\Gamma}

\journal{J. Quant. Spectrosc. Radiat. Transf.}

\begin{document}

\begin{frontmatter}

\title{A branchless single-precision Voigt approximation for spectral synthesis\tnoteref{am}}
\tnotetext[am]{Accepted manuscript of an article accepted for publication in the Journal of Quantitative Spectroscopy and Radiative Transfer. \copyright\ 2026 The Author. Published open access under the CC BY 4.0 license, \url{https://creativecommons.org/licenses/by/4.0/}.}

\author[mpia,uni]{Richard Hoppe\orcidlink{0000-0002-8451-6260}\corref{cor}}
\ead{hoppe@mpia.de}
\cortext[cor]{Corresponding author.}
\affiliation[mpia]{organization={Max-Planck-Institut für Astronomie},
                   addressline={Königstuhl 17},
                   postcode={69117},
                   city={Heidelberg},
                   country={Germany}}
\affiliation[uni]{organization={Ruprecht-Karls-Universität Heidelberg},
                  postcode={69117},
                  city={Heidelberg},
                  country={Germany}}

\begin{abstract}
The Voigt function is evaluated repeatedly in spectral synthesis, and several widely used implementations carry an absolute error floor. We construct QLG, a family of branchless, single-precision approximations, separable in the damping parameter $a$. QLG is named for its three ingredients: Lorentzian and Gaussian generators $L$ and $G$, and an envelope $q$ that multiplies every fitted correction and vanishes at both endpoints. The leading Lorentzian wing amplitude is fixed analytically. Three members reach observed maximum full-profile relative errors of $6.5\times10^{-3}$, $9.2\times10^{-4}$, and $7\times10^{-5}$ over $a\in[10^{-9},10^{9}]$ and $|v|\le10^{5}$. Extended-domain and randomized checks support these error levels. For the tested \textsc{siu}, \textsc{moog}, and Huml\'i\v{c}ek \texttt{w4} implementations, the relative error instead approaches $100\%$ towards small values of the damping parameter. We compare synthesis-weighted costs on an Apple M1 and an AMD EPYC 9554 using \texttt{gfortran 15}, and propagate errors through solar-dwarf, red-supergiant, and M-dwarf syntheses. The least expensive member costs $0.98$--$1.05$ times as much as the tabulated \textsc{Synspec} kernel on the M1 and $1.11$--$1.20$ times as much on the EPYC, with a maximum flux error smaller by a factor of $3.6$. The intermediate member costs less than every tested arithmetic kernel and holds the flux error below $10^{-3}$. The most accurate member costs $1.6$--$3.0$ times less than \texttt{w4} and gives $3.4$--$5.0$ times smaller maximum flux errors in the tested syntheses.
\end{abstract}


\end{frontmatter}

\section{Introduction}
\label{sec:intro}

Current and new-generation multiplexed surveys are producing large stellar spectroscopic data sets \citep{deJongAgertz2019,DaltonTrager2012,KollmeierZasowski2017,CirasuoloAfonso2014}. Deriving atmospheric parameters and chemical abundances from them requires repeated comparisons with synthetic spectra, often across extensive model grids or many fitting iterations, while the atomic and molecular line databases used in these calculations continue to grow \citep{GordonRothman2026,TennysonYurchenko2012,PasseggerSchweitzer2019,Birkby2018}.

The Voigt profile is evaluated repeatedly in the inner loops of line-by-line stellar and molecular spectral synthesis \citep{Sneden1973,ValentiPiskunov1996,Plez2012,WheelerAbruzzo2023}. A useful approximation must remain accurate in the damping wings because strong lines contribute opacity far below the profile peak. A small absolute-error floor can become a large fractional opacity error in these wings and can consequently alter equivalent widths and emergent flux \citep{Schreier2018}. We therefore adopt the maximum relative error over the full evaluated profile as the primary kernel-level metric and test its relevance through complete spectral syntheses.

Reducing the number of evaluated profile points is complementary to reducing the cost of each evaluation \citep{YurchenkoAmundsen2017,Min2017}. Here we hold the synthesis and the line-window criterion fixed and compare only the profile kernels. Branchless approximations to the Voigt function have been published before \citep{Kielkopf1973}. We introduce QLG, a family of branchless Voigt approximations constructed from Gaussian and Lorentzian generators, the two elementary functions from which every basis function of the expansion is built. The approximation is separable in the damping parameter $a$: every dependence on $a$ is confined to factors constant within a line, so the point loop, the sweep over one line window's wavelength points, evaluates functions of the frequency displacement alone. We select three members for three deployment cases, compare their profile accuracy and synthesis-weighted cost with published kernels in a common benchmark program, and measure how the kernel errors propagate through the emergent spectra.

\section{Approximation}
\label{sec:method}

We use the dimensionless Voigt function
\begin{equation}
  \K(a,v)
  = \mathrm{Re}\!\left[w(v+\mathrm{i}a)\right]
  = \frac{a}{\pi}\int_{-\infty}^{\infty}
    \frac{\mathrm{e}^{-t^{2}}}{a^{2}+(v-t)^{2}}\,\mathrm{d}t,
  \label{eq:voigt}
\end{equation}
where $w(z)=\mathrm{e}^{-z^{2}}\mathrm{erfc}(-\mathrm{i}z)$ is the Faddeeva function \citep{FaddeyevaTerentev1961}, $a$ is the damping parameter, and $v$ is the frequency displacement in Doppler-width units. The expansion is written on an approximate half-width $\hw(a)$. Two forms are used below. The first is the empirical approximation of \citet{Whiting1968} with the coefficients refined by \citet{Olivero1977},
\begin{equation}
  \hw(a) = 0.5346\,a + \sqrt{0.2166\,a^{2}+\ln 2}.
  \label{eq:hwhm}
\end{equation}
The second avoids the square root,
\begin{equation}
  \frac{1}{\hw(a)} = \frac{d(a)}{a\,d(a)+\sqrt{\ln 2}},
  \qquad d(a)=1+d_{1}a,
  \label{eq:hwrat}
\end{equation}
and its coefficient $d_{1}$ is fitted jointly with the expansion coefficients. Both forms satisfy $\hw(0)=\sqrt{\ln 2}$. Equation~(\ref{eq:hwrat}) also satisfies $\hw/a\rightarrow1$ as $a\rightarrow\infty$; the rounded coefficients in Eq.~(\ref{eq:hwhm}) instead give $\hw/a\rightarrow1.000003051$, leaving a small large-$a$ error (\ref{app:fitting}).

The scaled variables and the Gaussian and Lorentzian generators are
\begin{align}
  r &= \frac{a}{\hw}, \quad q = r(1-r), \quad x = 2r-1, \nonumber \\
  u &= \left(\frac{v}{\hw}\right)^{2}, \quad L = \frac{1}{1+u}, \quad G = 2^{-u}.
  \label{eq:LG}
\end{align}
The variable $r$ parameterizes the transition from the Gaussian limit at $r=0$ to the Lorentzian limit at $r=1$, while $L$ maps $|v|\in[0,\infty)$ onto $L\in(0,1]$ and converts the asymptotic condition at $|v|\rightarrow\infty$ into a boundary condition at $L=0$. We retain $G$ as a separate generator because the Gaussian limit is flat to all orders at $L=0$ and cannot be represented efficiently by a finite polynomial in $L$. The shifted variable $x$ is the argument of every fitted coefficient polynomial in place of $r$, because it keeps their coefficients one to two orders of magnitude smaller and so limits cancellation in \texttt{fp32} (\ref{app:fitting}).

We define the width-scaled profile
\begin{equation}
  F(r,u)=\frac{\hw(a)}{\sqrt{\pi}}\,\K(a,v),
  \label{eq:scaled_profile}
\end{equation}
which can be expressed in terms of $r$ and $u$. The Lorentzian far wing satisfies $\K(a,v)\sim a/(\sqrt{\pi}v^{2})$ as $|v|\rightarrow\infty$, which fixes the leading wing term without fitting: since $L\sim\hw^{2}/v^{2}$ in this limit, its required coefficient in $F$ is $r/\pi$. Multiplying every fitted correction by $q=r(1-r)$ preserves the Gaussian and Lorentzian profiles at the formal endpoints $r=0$ and $r=1$. The resulting family is named for its three ingredients, the two generators $L$ and $G$ and the envelope $q$ that multiplies every fitted departure from them, and we abbreviate it QLG. It has the form
\begin{align}
  F(r,u)
  &= \underbrace{\frac{r}{\pi}L
    + \sqrt{\frac{\ln 2}{\pi}}\,(1-r)G}_{\textstyle P(r,u)}
    + q\sum_i c_i(x)\,\phi_i(u),
  \label{eq:qlg_general}
\end{align}
where each coefficient $c_i(x)$ is a polynomial in $x$ and the basis functions $\phi_i$ are products $G^{j_i}L^{k_i}$ of the two generators. The approximation is linear in all unknown constants, which we determine jointly in one overdetermined linear least-squares fit on a relative-error objective over the sampled $(r,L)$ domain. The basis of each member was selected from a pool of candidate shapes; the fitting and the selection procedure are given in \ref{app:fitting}.

Two constraints act on the pool. The correction must vanish at $r=0$ and $r=1$, which $q$ enforces for any admissible $\phi_i$, and every correction basis function must decay faster than the leading Lorentzian term, which bars $G^{0}L^{k}$ with $k<2$. Both are imposed before fitting and hold for every member below.

The smallest useful member is compact enough to state here. QLG-toy uses three basis functions at $x$-degree zero, so its three coefficients are plain numbers:
\begin{equation}
  F(r,u) = \frac{r}{\pi}L + \sqrt{\frac{\ln 2}{\pi}}\,(1-r)G
    + q\left(c_1 L^{2} + c_2 G + c_3 G^{2}L^{2}\right),
  \label{eq:qlg3}
\end{equation}
with $c_1=1.413611936$, $c_2=-0.635779502$, and $c_3=-0.850801079$ on the half-width of Eq.~(\ref{eq:hwhm}). Both counts are minimal at its accuracy of $2.2\times10^{-2}$ (\ref{app:fitting}).

\begin{figure}[t]
\centering
\includegraphics[scale=0.7]{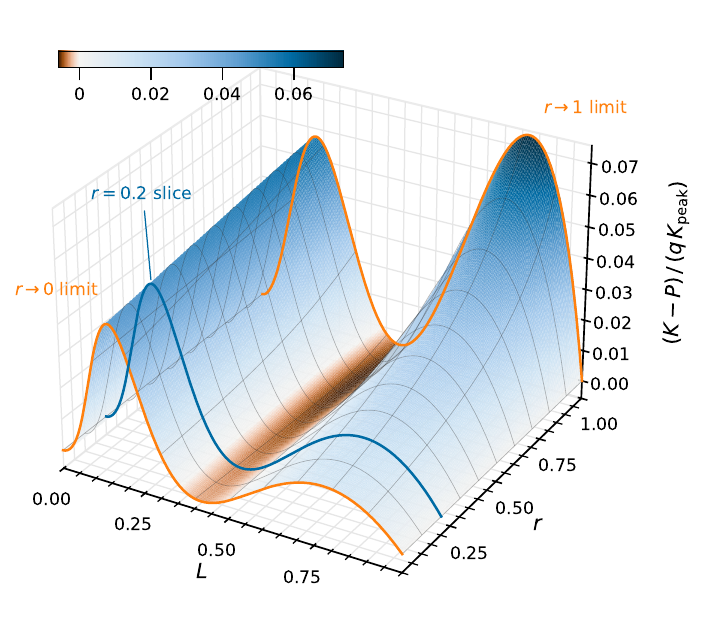}
\caption{The residual $[F-P]/q$ of Eq.~(\ref{eq:qlg_general}) that the fitted correction must represent, over the $(r,L)$ domain, normalized to the profile peak. The colour map is centred on zero, so its orange half covers the small negative range. A single slice at fixed $r$, marked here at $r=0.2$, approximates the residual at every $r$ to within the accuracy required of the lowest tier. That slice is the basis function $Z_{0.2}$ of Eq.~(\ref{eq:zshape}), which QLG-bot reads from a table.}
\label{fig:pseudo}
\end{figure}

The residual $[F-P]/q$ that Eq.~(\ref{eq:qlg_general}) has to represent depends only weakly on $r$ (Fig.~\ref{fig:pseudo}), so a single slice of it at fixed $r=r_{*}$,
\begin{equation}
  Z_{r_{*}}(u)=\frac{F(r_{*},u)-P(r_{*},u)}{q(r_{*})},
  \label{eq:zshape}
\end{equation}
is a function of $u$ alone and can be tabulated and used as a further basis function. An implementation that already reads $G$ from a table can read $Z_{r_{*}}$ at the same index, at a cost of one memory access and one multiply-add. QLG-bot uses $r_{*}=0.2$.

\label{sec:tiers}
We select one member for each of three deployment cases, and name it for its position in the family:

QLG-bot was selected to match the cost of the tabulated \textsc{Synspec} kernel on the M1, the least expensive comparison kernel on that processor. It uses three basis functions, $G$, $Z_{0.2}$, and $L^{2}$, with linear coefficients, and the half-width of Eq.~(\ref{eq:hwrat}) with $d_{1}=0.60602$. Six fitted constants.

QLG-mid is the least expensive member whose maximum relative error is below $10^{-3}$. It uses seven basis functions with linear coefficients on the half-width of Eq.~(\ref{eq:hwhm}). Fourteen fitted constants.

QLG-top is the least expensive member intended as a replacement for the Huml\'i\v{c}ek \texttt{w4} kernel. It uses thirteen basis functions of $x$-degree two to four on the half-width of Eq.~(\ref{eq:hwhm}). Fifty-six fitted constants.

QLG-mid and QLG-top evaluate $G$ arithmetically, through IEEE-754 exponent handling (\ref{app:coeff}); QLG-bot reads it from a table, with $Z_{0.2}$.

The coefficients of all three are listed in \ref{app:coeff}. Which half-width is the better of the two depends on the accuracy of the member that uses it (\ref{app:fitting}).

\section{Cost and profile accuracy}
\label{sec:results}

\begin{table*}[t]
\centering
\begin{threeparttable}
\caption{Profile accuracy and synthesis-weighted cost of the three QLG members and the comparison kernels. The cost columns give the elapsed time per evaluated point over the recorded line windows of the solar-dwarf, red-supergiant, and M-dwarf syntheses on an Apple M1 (AArch64) and on an AMD EPYC 9554 (x86-64), each compiled with \texttt{gfortran 15}. Maximum relative errors are over the validation domain $a\in[10^{-9},10^{9}]$, $|v|\le10^{5}$, and the flux column gives the largest normalized-flux error among the three syntheses. All values were measured on the compiled \texttt{fp32} kernels (\ref{app:method}).}
\label{tab:deployed}
\scriptsize
\setlength{\tabcolsep}{3.5pt}
\begin{tabular}{llrrrrrrrr}
\toprule
kernel & algorithm &
\multicolumn{3}{c}{Apple M1 (ns point$^{-1}$)} &
\multicolumn{3}{c}{AMD EPYC 9554 (ns point$^{-1}$)} &
max rel. error & flux \\
\cmidrule(lr){3-5}\cmidrule(lr){6-8}
 & & sun & giant & M dwarf & sun & giant & M dwarf & full profile & max $|\delta F/F|$ \\
\midrule

QLG-bot
 & branchless tabulated
 & 2.05 & 2.55 & 3.38 & 2.25 & 2.98 & 4.02
 & $6.5\times10^{-3}$
 & $4.2\times10^{-3}$ \\

QLG-mid
 & branchless closed form
 & 2.62 & 3.58 & 5.30 & 2.59 & 4.15 & 6.67
 & $9.2\times10^{-4}$
 & $7.5\times10^{-4}$ \\

QLG-top
 & branchless closed form
 & 3.36 & 4.86 & 7.54 & 3.47 & 5.70 & 9.30
 & $7\times10^{-5}$
 & $1.2\times10^{-5}$ \\

\midrule

\textsc{Synspec}, \textsc{Tlusty}
 & Landolt--B\"ornstein, tabulated
 & 2.09 & 2.44 & 3.21 & 2.03 & 2.49 & 3.48
 & $1.3\times10^{-1}$\tnote{b}
 & $1.5\times10^{-2}$ \\

\textsc{siu}\tnote{c}
 & Landolt--B\"ornstein
 & 3.91 & 5.63 & 5.98 & 9.78 & 6.84 & 9.76
 & $3.5\times10^{-1}$\tnote{a}
 & $2.5\times10^{-2}$ \\

\textsc{moog}, \textsc{Korg}\tnote{c}
 & Landolt--B\"ornstein
 & 4.51 & 7.07 & 7.04 & 10.33 & 8.36 & 10.78
 & $3.7\times10^{-1}$\tnote{a}
 & $2.5\times10^{-2}$ \\

\textsc{SMElib}
 & Huml\'i\v{c}ek \texttt{w4}
 & 6.23 & 10.32 & 11.89 & 7.79 & 12.71 & 15.41
 & $3.7\times10^{-3}$\tnote{a}
 & $5.7\times10^{-5}$ \\

\textsc{Turbospectrum}
 & Huml\'i\v{c}ek \texttt{w4}
 & 6.79 & 12.38 & 15.19 & 8.66 & 17.29 & 22.16
 & $3.7\times10^{-3}$\tnote{a}
 & $5.7\times10^{-5}$ \\

\midrule
HAW \citep{Hui1978}
 & $6/7$ rational, clamped
 & 3.49 & 4.19 & 5.85 & 3.46 & 5.39 & 8.11
 & $5.1\times10^{3}$\tnote{a}
 & $9.0\times10^{-1}$ \\
\bottomrule
\end{tabular}
\begin{tablenotes}\scriptsize
\item All kernels except HAW return finite, non-negative values on the validation grid. HAW returns negative inner-wing values and nonfinite values at $12.9\%$ of the grid points. Its profile-error entry is the maximum over finite outputs of the unclamped kernel; its cost and flux columns use the kernel clamped at zero (\ref{app:method}).
\item[a] Attained at the lower boundary $a=10^{-9}$; the value depends on where that boundary is placed (Section~\ref{sec:accuracy}).
\item[b] Attained at $a=1.4$, interior to the domain, at a branch test of the kernel's own piecewise construction (\ref{app:method}).
\item[c] On the EPYC, the compiler vectorises the branched point loop and evaluates $\exp(-v^{2})$ for every point, including far-wing points that the branch discards, where the vectorised library exponential is slow (\ref{app:method}).
\end{tablenotes}
\end{threeparttable}
\end{table*}

The comparison kernels are representative implementations of the named codes' formulas rather than the codes themselves, transcribed into a common benchmark program and timed under identical conditions; profile errors were evaluated on the compiled kernels against \texttt{scipy.special.wofz} in \texttt{fp64} \citep{VirtanenGommers2020}. \ref{app:method} gives the source of each transcription, the compilation and timing procedure, and the validation grid. The measurements below establish performance only for the two tested CPUs, the compiler, and the kernel implementations.

\subsection{Synthesis-weighted cost}

We recorded the line windows evaluated in three complete syntheses: a solar dwarf ($T_{\rm eff}=5777$\,K and $\log g=4.4$), a red supergiant ($T_{\rm eff}=4000$\,K and $\log g=1.0$), and an M dwarf ($T_{\rm eff}=4000$\,K and $\log g=4.5$), all from the \textsc{marcs} grid \citep{GustafssonEdvardsson2008}. Each synthesis assumed solar metallicity and covered $3000$--$9000$\,\AA{} at $R=10^{5}$. Each line window is terminated when its opacity falls below either $10^{-4}$ of the local continuum opacity or below $10^{-8}$ of the total accumulated opacity including blending lines. The M-dwarf workload is concentrated at narrower windows than those of the other two atmospheres and therefore provides the least favourable case for kernels with significant per-line setup work.

We then replayed the recorded stream of $(n_w,a,\Delta v)$ triples through each kernel, retaining wide windows in full and subsampling narrow windows at a fixed stride with statistical weights, so that every kernel met the same distribution of setup and point-evaluation work. The reported cost is the synthesis-weighted elapsed time per evaluated point (\ref{app:method}).

\subsection{Computational cost}

On the M1, QLG-bot requires $2.05$--$3.38$\,ns per evaluated point across the three atmospheres, QLG-mid $2.62$--$5.30$\,ns, and QLG-top $3.36$--$7.54$\,ns (Table~\ref{tab:deployed}). QLG-bot costs $0.98$--$1.05$ times as much as the tabulated \textsc{Synspec} kernel, and no kernel we tested is less expensive than either. QLG-mid is less expensive than every tested kernel that evaluates the profile arithmetically, on all three atmospheres, by factors of $1.10$--$3.46$. QLG-top is less expensive than both \texttt{w4} implementations by factors of $1.58$--$2.55$; against \textsc{siu} and \textsc{moog} it is less expensive on the solar dwarf and the red supergiant and more expensive on the M dwarf, whose narrow windows make per-line setup comparatively more important.

On the EPYC, QLG-bot costs $1.11$--$1.20$ times as much as \textsc{Synspec}. QLG-mid is less expensive than every tested arithmetic kernel by factors of $1.22$--$4.17$, and QLG-top is less expensive than both \texttt{w4} implementations by factors of $1.66$--$3.03$.

The cost of a member is set by its arrangement and not by its coefficient values, so a candidate can be priced before it is fitted and the search restricted to arrangements inside a budget (\ref{app:fitting}).

\subsection{Profile accuracy}
\label{sec:accuracy}

An omitted Gaussian tail can make the maximum relative error sensitive to the domain boundary. The \textsc{siu} and \textsc{moog} implementations of the Landolt--B\"ornstein formula switch to its asymptotic Lorentzian wing at $|v|=5$ for $a\le0.2$ and omit the Gaussian term there, an absolute error of up to $\mathrm{e}^{-25}\simeq1.4\times10^{-11}$ of the profile peak. As $a\rightarrow0$ the Lorentzian component $a/(\sqrt{\pi}v^{2})$ vanishes and only this Gaussian tail remains, so their sampled maximum relative error is attained at the lower boundary $a=10^{-9}$ and $|v|\simeq5$, and rises from $35$--$37\%$ there to nearly $100\%$ by $a=10^{-12}$. Both \texttt{w4} implementations also attain theirs at the lower boundary; their relative error initially grows approximately as $1/a$, exceeds $0.5$ below $a\simeq3\times10^{-12}$, and then approaches $100\%$. The tabulated \textsc{Synspec} kernel evaluates the same Landolt--B\"ornstein formula but switches to the asymptotic wing only beyond $|v|=10$, where the omitted Gaussian term is below $10^{-43}$; its relative error remains at a few per cent in the small-$a$ checks down to $10^{-18}$ (\ref{app:method}).

QLG retains the Gaussian term rather than dropping it at a fixed displacement. The factor $q=r(1-r)$ removes the fitted correction at $a=0$, recovering the Gaussian profile, and every correction basis function decays at least as fast as $L^{2}$, preserving the leading Lorentzian wing. The observed maxima of the three members lie in the interior of the validation domain, at $a=0.044$, $1.57$, and $0.173$. Extending the sampled domain to $a=10^{-18}$ and $|v|=10^{8}$ leaves these error levels unchanged, and $3\times10^{7}$ randomly drawn $(a,v)$ pairs reproduce the QLG-bot and QLG-mid maxima to within $0.2\%$. The QLG-top maximum is set by \texttt{fp32} rounding in the setup and was located by evaluating every \texttt{fp32} value of $a$ in the interval that contains it (\ref{app:method}).

Within the tabulated domain, QLG-mid reaches $9.2\times10^{-4}$ and QLG-top $7\times10^{-5}$, against $3.5$--$3.7\times10^{-1}$ for \textsc{siu} and \textsc{moog} and $1.3\times10^{-1}$ for the tabulated \textsc{Synspec} kernel, three orders of magnitude smaller for QLG-top and two for QLG-mid. A comparison against \texttt{w4}'s $3.7\times10^{-3}$ is not decisive on this metric, because that value is set by the domain boundary rather than by \texttt{w4}'s typical error; Section~\ref{sec:endtoend} settles it with the emergent-flux propagation. Above $a=20$, \texttt{w4} is the more accurate of the two, where QLG-top reaches $3.2\times10^{-6}$.

Over the validation domain all three members are non-negative and decrease monotonically with $|v|$ to within \texttt{fp32} rounding (\ref{app:fitting}). HAW \citep{Hui1978} instead becomes negative in the inner wing below $a\simeq2.4\times10^{-6}$. A synthesis using it must therefore clamp the profile, with the consequences quantified in Section~\ref{sec:endtoend}.

\section{Propagation to emergent spectra}
\label{sec:endtoend}

We substituted each selected kernel into the \texttt{OpenMULTI3D} synthesis code (Hoppe et al., subm.) while holding the atmospheres, line data, opacity sources, spectral sampling, and accumulated-opacity window criterion fixed. The calculations cover the same three atmospheres and the same wavelength range as the timing measurements. Each spectrum was compared with a synthesis using the \texttt{fp64} Huml\'i\v{c}ek--Weideman-plus-asymptotic reference kernel \citep{Schreier2011}, and the reported statistic is the maximum fractional flux residual, $\max_{\lambda}|F_k(\lambda)-F_{\rm ref}(\lambda)|/F_{\rm ref}(\lambda)$, listed in Table~\ref{tab:deployed}.

QLG-mid and QLG-top hold every wavelength point of all three spectra below $10^{-3}$, at maxima of $7.5\times10^{-4}$ and $1.2\times10^{-5}$. QLG-bot exceeds $10^{-3}$ at $1.4$--$2.2\%$ of wavelength points at a maximum of $4.2\times10^{-3}$, against $3.0$--$5.3\%$ and $1.5\times10^{-2}$ for the tabulated \textsc{Synspec} kernel of comparable cost. Distributions for all three members are given in \ref{app:method}.

The sampled maximum full-profile relative errors, restricted to finite outputs for HAW, order the synthesized kernels as QLG-top, QLG-mid, \texttt{w4}, QLG-bot, \textsc{Synspec}, \textsc{siu} and \textsc{moog}, and HAW. The emergent-flux errors follow this ordering with one exception: \texttt{w4} produces a smaller flux error than QLG-mid despite a larger full-profile error, because its maximum is attained at damping parameters four orders of magnitude below the smallest value the three syntheses sample. Restricted to the Doppler core, its maximum relative error is $9.6\times10^{-5}$ and the ordering is reproduced.

The broad damping wings of Ca\,\textsc{ii} H\&K show where a kernel with a small absolute error can still fail. HAW has a maximum absolute error of $4.5\times10^{-6}$ of the profile peak, but its relative error grows in the weak-profile wing and the approximation eventually becomes negative. Clamping these values to zero during synthesis prevents negative opacity but removes the affected wing contribution, and the resulting profile is unphysical: the broad wings of Ca\,\textsc{ii}\,K are visibly suppressed and the line is far too weak (Fig.~\ref{fig:cahk}), and the clamped synthesis reaches a flux error of $9.0\times10^{-1}$.

\section{Conclusions}
\label{sec:conclusions}

QLG combines a form separable in $a$ and a branchless point loop with analytically fixed endpoint profiles and Lorentzian far-wing amplitude. The observed maximum relative errors lie in the interior of the validation domain and are supported by extended-domain and randomized checks. For the tested \textsc{siu} and \textsc{moog} implementations of the Landolt--B\"ornstein formula and for \texttt{w4}, the corresponding figure approaches $100\%$ as the domain is extended towards small $a$.

The three members cover the cost range spanned by the tested production kernels. QLG-bot costs $0.98$--$1.05$ times as much as the tabulated \textsc{Synspec} kernel on the M1 and $1.11$--$1.20$ times as much on the EPYC, at a maximum flux error smaller by a factor of $3.6$. QLG-mid is less expensive than every tested kernel that evaluates the profile arithmetically and holds the flux error of all three spectra below $10^{-3}$. QLG-top is less expensive than both \texttt{w4} implementations by factors of $1.58$--$3.03$ on the two processors and reduces the maximum flux error from $5.7\times10^{-5}$ to $1.2\times10^{-5}$, so for the three syntheses examined here a code currently using \texttt{w4} can adopt it without an accuracy cost.

Maximum relative error over the full profile is inexpensive enough to serve as a screening metric and reproduced the emergent-flux ordering, with the one exception noted above. Peak-normalized absolute error did not identify kernels that suppress astrophysically important damping wings.

The present benchmarks establish CPU performance on one AArch64 and one x86-64 processor with one compiler, and error propagation in stellar synthesis; performance on GPUs and suitability for molecularly dominated or exoplanet spectra remain to be tested.

\section*{Declaration of competing interest}

The author declares no known competing financial interests or personal relationships that could have influenced the work reported in this paper.

\section*{Data availability}

The Fortran module implementing the three members and the reference kernel (\texttt{qlg\_mod.f90}), the benchmark replay program, and the recorded line-window streams of the three syntheses are supplied as supplementary material.

\section*{Declaration of generative AI and AI-assisted technologies in the writing process}

During the preparation of this work the author used Claude Code and Chat AI, a service providing access to large language models hosted by GWDG \cite{doosthosseini_saia_2026}, to write scripts for fitting and timing the QLG members and to draft text. After using these tools, the author reviewed and edited the content as needed and takes full responsibility for the content of the published article.

\section*{Acknowledgements}

RH thanks A.~E.~Lynas-Gray and an anonymous referee for their comments, which improved the manuscript. RH is supported through the Lise Meitner grant from the Max Planck Society and through the ERC grant of the European Research Council (ERC) under the European Union's Horizon 2020 research and innovation programme (Grant agreement No. 949173). This work used NumPy \cite{Harris2020}, SciPy \cite{VirtanenGommers2020}, and Matplotlib \cite{Hunter2007}.

\bibliographystyle{elsarticle-num-names}
\bibliography{refs}

\appendix
\counterwithin*{figure}{section}
\counterwithin*{table}{section}
\input{qlg_appendices}

\end{document}

%% file: qlg_appendices.tex
\section{Coefficients and implementation}
\label{app:coeff}

For every member,
\begin{equation}
  K(a,v)=\frac{\sqrt{\pi}}{\Gamma(a)}F(r,u),
\end{equation}
with $\Gamma$, $r$, $q$, $x$, $u$, $L$, and $G$ defined in Eqs.~(\ref{eq:hwhm})--(\ref{eq:LG}).
The correction in Eq.~(\ref{eq:qlg_general}) is written
\begin{equation}
  q\sum_i c_i(x)\,\phi_i(u),
  \qquad
  c_i(x)=\sum_{m=0}^{d_i}c_{im}x^{m} ,
  \label{eq:coefficient_convention}
\end{equation}
and Table~\ref{tab:qlg_coeff} lists the coefficients in ascending powers of $x$. Blank entries are zero. QLG-bot uses the half-width of Eq.~(\ref{eq:hwrat}) with $d_{1}=0.6060220921$; QLG-mid and QLG-top use Eq.~(\ref{eq:hwhm}).

\begin{table*}[t]
\centering
\caption{Coefficients of the three QLG members in the convention of Eq.~(\ref{eq:coefficient_convention}). All three are evaluated in \texttt{fp32}. The basis function $Z_{0.2}$ is defined by Eq.~(\ref{eq:zshape}) and tabulated as described in \ref{app:tables}.}
\label{tab:qlg_coeff}
\scriptsize
\setlength{\tabcolsep}{4pt}
\begin{tabular}{lrrrrr}
\toprule
basis & $c_{i0}$ & $c_{i1}$ & $c_{i2}$ & $c_{i3}$ & $c_{i4}$ \\
\midrule
\multicolumn{6}{l}{\textbf{QLG-bot} (6 constants, half-width Eq.~\ref{eq:hwrat})} \\
$G$        & $-0.06492439319$ & $-0.1101932628$ & & & \\
$Z_{0.2}$  & $ 0.3849419593$  & $-0.1056713528$ & & & \\
$L^{2}$    & $ 0.04054415746$ & $ 0.1624102925$ & & & \\

\midrule
\multicolumn{6}{l}{\textbf{QLG-mid} (14 constants, half-width Eq.~\ref{eq:hwhm})} \\
$G$        & $-0.789538867$   & $ 0.3148034058$   & & & \\
$GL^{-1}$  & $-0.04265367129$ & $ 0.09756947369$  & & & \\
$L^{2}$    & $ 1.18707601$    & $ 0.1145307711$   & & & \\
$GL^{2}$   & $-5.720195813$   & $ 11.42426912$    & & & \\
$GL^{3}$   & $-10.74044061$   & $ 15.57648767$    & & & \\
$L^{4}$    & $ 32.14587012$   & $-58.42788171$    & & & \\
$GL^{4}$   & $-16.07919633$   & $ 30.8985669$     & & & \\

\midrule
\multicolumn{6}{l}{\textbf{QLG-top} (56 constants, half-width Eq.~\ref{eq:hwhm})} \\
$G$        & $-0.8646222277$  & $ 1.196227073$   & $-1.946778657$    & $ 0.8700489847$    & \\
$L^{2}$    & $ 1.141897161$   & $ 0.1391619211$  & $ 0.001768314879$ & & \\
$G^{2}L^{2}$ & $-5.386293023$ & $-2.67138434$    & $ 39.87611422$    & $-27.97250459$     & $ 1.403038488$ \\
$L^{3}$    & $ 2.279531261$   & $-2.901926085$   & $-0.3081531094$   & $ 0.2136156197$    & $-0.007055088033$ \\
$G^{2}$    & $-0.5365843159$  & $ 0.8589181727$  & $ 2.652786464$    & $-3.128877765$     & $ 0.3580500169$ \\
$GL$       & $-0.8064421971$  & $-3.389981855$   & $ 17.56308532$    & $-11.8836514$      & $ 0.2252746071$ \\
$L^{4}$    & $ 6.469699634$   & $-0.1679614655$  & $-14.35427448$    & $-5.158498665$     & \\
$GL^{4}$   & $-4.503441977$   & $ 77.21445254$   & $-117.9185115$    & & \\
$G^{2}L$   & $-1.964906322$   & $ 0.6071469184$  & $ 11.63689108$    & $-8.97934044$      & $-0.1078078006$ \\
$GL^{2}$   & $ 3.915934761$   & $ 11.83234477$   & $-70.72115163$    & $ 48.13945138$     & $-1.71403342$ \\
$GL^{5}$   & $ 0.7967260847$  & $ 111.5775369$   & $-191.4626908$    & & \\
$L^{5}$    & $-0.9859997764$  & $-223.0961222$   & $ 380.4154228$    & $ 0.8698176889$    & $-0.1589890977$ \\
$GL^{3}$   & $ 0.4049704817$  & $ 28.79962278$   & $-55.43619176$    & $ 7.028235913$     & \\
\bottomrule
\end{tabular}
\end{table*}

\subsection{Gaussian evaluation}

QLG-mid and QLG-top evaluate $G=2^{-u}$ by writing $u=n+f$, where $n=\lfloor u\rfloor$ and
$0\le f<1$. Thus
\begin{equation}
  G=2^{-n}p_2(f),
  \qquad
  p_2(f)=\sum_{k=0}^{5}b_k f^k ,
  \label{eq:exp2_implementation}
\end{equation}
and the factor $2^{-n}$ is formed through the IEEE-754 exponent field. The polynomial
coefficients are
\begin{equation}
\begin{aligned}
 b_0 &= \phantom{-}0.9999999404,\\
 b_1 &=            -0.6931422353,\\
 b_2 &= \phantom{-}0.2401721776,\\
 b_3 &=            -0.05527971312,\\
 b_4 &= \phantom{-}0.009188604541,\\
 b_5 &=            -0.0009387863684.
\end{aligned}
\end{equation}
Before conversion to an integer, $u$ is replaced by $\min(u,127)$. The endpoint therefore produces zero in \texttt{fp32}, preventing a residual Gaussian floor in the Lorentzian wing. The maximum relative error of the rounded polynomial over $0\le f<1$ is $1.2\times10^{-7}$, and its contribution to the error of QLG-mid and QLG-top is below $1\%$ of their respective maxima.

\subsection{Tabulated generators}
\label{app:tables}

QLG-bot reads both of its transcendental factors from tables indexed on $\xi=|v|/\Gamma=\sqrt{u}$, with $800$ intervals per unit $\xi$ over $0\le\xi\le11.27$ and nearest-node lookup. The first table holds $G=2^{-u}$. The second holds $Z_{0.2}$ of Eq.~(\ref{eq:zshape}) divided by its known $L^{2}$ wing weight,
\begin{equation}
  W_{0.2} = Z_{0.2}\,(1+u)^{2} = Z_{0.2}/L^{2} ,
\end{equation}
with the factor $L^{2}$ restored arithmetically after the read. Both tables are indexed by the same snapped integer, so the second costs one memory access and one multiply-add. Entries of the $G$ table with $u>126$ lie below the smallest normal \texttt{fp32} number and are stored as zero, the far-wing limit of $G$; on x86-64 processors that do not flush subnormal operands to zero, a subnormal entry read at the clamped index would incur a floating-point assist at every far-wing point.

The table sizes, initialization routine, evaluation order, and pre-scaled coefficients are contained in \texttt{qlg\_mod.f90}. The tabulated entry point requires one call to \texttt{qlg\_init} before its first use.

\section{Fitting and basis selection}
\label{app:fitting}

Only the correction term of Eq.~(\ref{eq:qlg_general}) is fitted; the two endpoint terms $(r/\pi)L$ and $\sqrt{\ln2/\pi}\,(1-r)G$ are fixed analytically and subtracted from the reference profile beforehand. No fitted correction coefficient can alter the profiles at $r=0$ and $r=1$, or the leading far-wing amplitude. The objective is the relative residual,
\begin{equation}
  \min_{\{c_{im}\}}\;
  \left\|
  \frac{q\sum_i\sum_m c_{im}x^{m}\phi_i(u)
        -\left[F-\frac{r}{\pi}L-\sqrt{\tfrac{\ln2}{\pi}}(1-r)G\right]}
       {F}
  \right\|,
\end{equation}
evaluated on a grid in $r$ and $L$ that resolves both endpoints, where the correction envelope $q$ vanishes, and the far wing logarithmically in $L$, which a uniform grid would leave almost unsampled. Reference values were obtained in \texttt{fp64} from \texttt{scipy.special.voigt\_profile}. The design matrix is column-normalized and solved by linear least squares, then polished by an iteratively reweighted least-squares scheme that converges toward the equioscillation minimax solution and reduces the maximum relative error by $25$--$30\%$ relative to the plain least-squares solution used for basis selection.

The candidate pool comprised $G^jL^k$ with $j=0$ and $2\le k\le8$, $j=1$ and $0\le k\le5$, $j=2$ and $0\le k\le4$, $j=3$ and $0\le k\le2$, together with $GL^{-1}$ and, for the tabulated arm, the slice $Z_{r_*}$ of Eq.~(\ref{eq:zshape}); $L^{1}$ is reserved for the analytically fixed far-wing term and never enters a fitted basis. A candidate member is a subset of the pool together with an $x$-degree $d_i$ assigned to each selected basis function. An exhaustive search over the pool establishes that both counts of QLG-toy, Eq.~(\ref{eq:qlg3}), are minimal at its accuracy of $2.2\times10^{-2}$: no two-constant set reaches better than $1.1\times10^{-1}$, and no two-basis-function set better than $6.5\times10^{-2}$ at any $x$-degree.

Selection proceeded within a cost budget rather than a budget on the number of constants: the instruction stream of a member is set by the number of powers of $L$ in its Horner chain, the number of table reads, the degree of $d(a)$, and the choice between an arithmetic and a tabulated $G$, so the cost of an arrangement could be approximated before any member using it was fitted and only arrangements inside its budget were fitted (Section~\ref{sec:results}).

\subsection{Choice of half-width}

Equations~(\ref{eq:hwhm}) and~(\ref{eq:hwrat}) are different one-parameter families and neither contains the other, so the better of the two depends on the accuracy of the member that uses it. Fitted jointly with the expansion coefficients at the accuracy of QLG-bot, Eq.~(\ref{eq:hwrat}) absorbs part of the shape error and reduces the maximum relative error by $5$--$12\%$. At the accuracy of QLG-mid the ordering reverses and Eq.~(\ref{eq:hwhm}) is better by a factor of $1.5$--$1.7$, and at that of QLG-top by a factor of $2.2$--$3.3$; a degree-two or degree-three $d(a)$, fitted at both accuracies, did not change this.

Eq.~(\ref{eq:hwhm}) carries its own asymptotic floor: its constants satisfy $0.5346+\sqrt{0.2166}=1.0000031$, so $\Gamma/a$ approaches $1.0000031$ rather than unity, and a relative error of $3.1\times10^{-6}$ survives at arbitrarily large $a$. This is below the maximum of every member listed here and is a consideration only for a tier below $\sim2\times10^{-5}$.

\subsection{Validation, precision, and profile properties}

Errors quoted in this work were measured on the compiled kernels (\ref{app:method}) and not on the fit.

Near the Gaussian--Lorentzian transition the correction sums signed terms that individually exceed their sum, so a rounding error is amplified by
\begin{equation}
  A=\max_{r,L}\;\frac{\sum_i|\text{term}_i|}{|F|},
\end{equation}
which is $2.9$ for QLG-toy and $272$ for a member of QLG-top's depth. Evaluating the coefficient polynomials in $x=2r-1$ rather than in $r$ keeps the \texttt{fp32} penalty below this bound: measured against an \texttt{fp64} evaluation through the same arithmetic, the single-precision loss is $1.6\times10^{-6}$ for QLG-bot and $3.0\times10^{-6}$ for QLG-mid, below $1\%$ of their maximum errors. For QLG-top the rounding of the setup differs between adjacent \texttt{fp32} values of $a$ and sets its maximum error (\ref{app:method}).

Over the validation domain every member is non-negative and decreases monotonically with $|v|$ to within rounding. Evaluated from the coefficients in \texttt{fp64} the largest upward step is at the machine epsilon; in the deployed \texttt{fp32} kernels it is $1$--$6\times10^{-7}$ of the local profile value.

\section{Benchmark methodology}
\label{app:method}

\begin{table*}[t]
\centering
\caption{Algorithms represented by the benchmark implementations, and the routine each transcription follows.}
\label{tab:kernel_sources}
\footnotesize
\setlength{\tabcolsep}{2pt}
\begin{tabular}{ll>{\raggedright\arraybackslash}p{0.20\textwidth}>{\raggedright\arraybackslash}p{0.23\textwidth}}
\toprule
kernel & formula & source followed & transcription \\
\midrule
QLG-bot
  & Eq.~(\ref{eq:qlg_general}), tabulated
  & this work
  & supplementary \texttt{qlg\_mod.f90} \\
QLG-mid, QLG-top
  & Eq.~(\ref{eq:qlg_general}), arithmetic
  & this work
  & supplementary \texttt{qlg\_mod.f90} \\
\textsc{siu}
  & Landolt--B\"ornstein \citep{Traving1960,Voigt1965}
  & \texttt{voigt.f90} (\textsc{siu})
  & real arithmetic, as in the source \\
\textsc{moog}, \textsc{Korg}
  & Landolt--B\"ornstein \citep{Traving1960,Voigt1965}
  & \texttt{Voigt.f} (\textsc{moog17scat})
  & independent transcription, same tree \\
\textsc{Synspec}, \textsc{Tlusty}
  & Landolt--B\"ornstein, tabulated \citep{Traving1960,Voigt1965}
  & \texttt{VOIGTK} (\texttt{synspec54.f})
  & nearest stored abscissa value \\
\textsc{SMElib}
  & Huml\'i\v{c}ek \texttt{w4} \citep{Humlicek1982}
  & \textsc{SMElib}/\textsc{sme}
  & complex division expanded into reals \\
\textsc{Turbospectrum}
  & Huml\'i\v{c}ek \texttt{w4} \citep{Humlicek1982}
  & \texttt{newvoigt.f} (v14.1--v20)
  & single-precision complex arithmetic \\
HAW
  & $6/7$ rational \citep{Hui1978}
  & \citet{Hui1978}
  & single-precision real arithmetic \\
\bottomrule
\end{tabular}
\end{table*}

\begin{table}[tbp]
\centering
\caption{Distribution of $|\delta F/F|$ against the \texttt{fp64} reference over
$3000$--$9000$\,\AA{} at $R=10^{5}$: the maximum, the $99$th percentile, and the fraction of wavelength points exceeding $10^{-3}$, for each of the three atmospheres.}
\label{tab:fluxdist}
\footnotesize
\setlength{\tabcolsep}{4pt}
\begin{tabular}{llrrr}
\toprule
kernel & atmosphere & max & $p_{99}$ & $>10^{-3}$ \\
\midrule
QLG-bot & sun     & $3.6\times10^{-3}$ & $1.3\times10^{-3}$ & $2.16\%$ \\
        & giant   & $4.2\times10^{-3}$ & $1.4\times10^{-3}$ & $1.67\%$ \\
        & M dwarf & $3.1\times10^{-3}$ & $1.1\times10^{-3}$ & $1.43\%$ \\
\midrule
QLG-mid & sun     & $5.9\times10^{-4}$ & $2.5\times10^{-4}$ & $0.00\%$ \\
        & giant   & $7.5\times10^{-4}$ & $3.3\times10^{-4}$ & $0.00\%$ \\
        & M dwarf & $5.3\times10^{-4}$ & $2.3\times10^{-4}$ & $0.00\%$ \\
\midrule
QLG-top & sun     & $1.1\times10^{-5}$ & $4.7\times10^{-6}$ & $0.00\%$ \\
        & giant   & $1.2\times10^{-5}$ & $5.6\times10^{-6}$ & $0.00\%$ \\
        & M dwarf & $9.3\times10^{-6}$ & $4.1\times10^{-6}$ & $0.00\%$ \\
\midrule
\textsc{Synspec}
        & sun     & $1.5\times10^{-2}$ & $2.2\times10^{-3}$ & $2.97\%$ \\
        & giant   & $1.2\times10^{-2}$ & $2.5\times10^{-3}$ & $3.59\%$ \\
        & M dwarf & $1.3\times10^{-2}$ & $2.5\times10^{-3}$ & $5.28\%$ \\
\midrule
\texttt{w4}
        & sun     & $3.6\times10^{-5}$ & $1.4\times10^{-5}$ & $0.00\%$ \\
        & giant   & $5.7\times10^{-5}$ & $2.2\times10^{-5}$ & $0.00\%$ \\
        & M dwarf & $4.6\times10^{-5}$ & $1.4\times10^{-5}$ & $0.00\%$ \\
\bottomrule
\end{tabular}
\end{table}

\subsection{Kernel implementations}

Table~\ref{tab:kernel_sources} identifies the formulas represented in the benchmark and the source each transcription follows. The comparison routines are representative implementations of those formulas, transcribed into a common Fortran interface in which $a$ is fixed for one line window and $v_f=v_0+(f-1)\Delta v$; they are not the original codes, and the reported costs are therefore costs of the formulas as implemented here rather than timings of the named codes. Per-line setup is included in the elapsed time.

The maximum relative errors of Table~\ref{tab:deployed} differ in character between the three implementations of the Landolt--B\"ornstein formula. \textsc{siu} and \textsc{moog} switch to the asymptotic Lorentzian wing at $|v|=5$ for $a\le0.2$ and omit the Gaussian term, an absolute error of up to $1.4\times10^{-11}$, so their sampled maximum relative error lies on the lower boundary of the validation domain and approaches $100\%$ as that boundary is lowered. The \textsc{Synspec} kernel makes the same switch only beyond $|v|=10$, where the omitted term is below $10^{-43}$. Its dominant small-$a$ error in the tested range is instead the snapping of $v$ to the nearest node of its pretabulated Harris functions, giving relative errors of about $3\times10^{-2}$ in checks down to $a=10^{-18}$. Its tabulated maximum of $1.3\times10^{-1}$ is attained at $a=1.4$, at a branch test of the kernel's own piecewise construction.

\subsection{Validation grid}

Profile errors were calculated as
\begin{equation}
  \epsilon_{\mathrm{rel}}(a,v)
  =
  \left|
  \frac{K_{\mathrm{test}}(a,v)}
       {K_{\mathrm{ref}}(a,v)}-1
  \right|
\end{equation}
on the values the compiled kernels return, so that the tabulated errors include the arithmetic of the deployed implementation and not only that of the fit. The validation domain was $a\in[10^{-9},10^9]$ and $|v|\le10^5$, sampled at $5217$ values of $a$ and $11\,062$ of $|v|$. The $a$ values comprise a logarithmic scan of the full domain together with $4000$ additional values over $0.005\le a\le5$ and selected model-selection and branch-boundary values; the $|v|$ values comprise a logarithmic scan together with the two uniform abscissae $v=0.125k$ and $v=0.0021+0.1234567k$, the second of which is incommensurate with the tabulated members' node spacing. The reference values were calculated in \texttt{fp64} using \texttt{scipy.special.wofz} \citep{VirtanenGommers2020}. The independent reference routine supplied in \texttt{qlg\_mod.f90} combines a Weideman rational approximation with a large-$|z|$ asymptotic expansion \citep{Schreier2011} and was used as a cross-check.

HAW returned nonfinite values (NaN or infinity) at $12.9\%$ of the grid points; these were counted separately and excluded from its finite-output maximum in Table~\ref{tab:deployed}. All other kernels returned finite values throughout the grid. HAW's profile-error entry describes the unclamped evaluation; its timing and synthesis results use zero-clamped outputs.

The coefficients were fitted on a grid in $(r,L)$ (\ref{app:fitting}) that is not aligned with the validation grid in $(a,v)$. Candidate members were ranked on $13$ values of $a$ between $10^{-8}$ and $86$, combined with the two uniform abscissae up to $|v|=500$, and this selection grid is a subset of the validation grid. The compiled kernels were therefore also evaluated at $3\times10^{7}$ randomly drawn pairs, $10^{7}$ in each of three samples: $a$ and $|v|$ log-uniform on $[10^{-9},10^9]$ and $[10^{-4},10^5]$, respectively; $a$ log-uniform on $[10^{-9},10^9]$ and $|v|$ uniform on $[0,500]$; and $a$ log-uniform on $[0.005,5]$ with $|v|$ uniform on $[0,50]$, which contains the observed maxima of all three members. The largest relative errors over the random samples are $6.50\times10^{-3}$, $9.218\times10^{-4}$, and $5.5\times10^{-5}$ for QLG-bot, QLG-mid, and QLG-top, against $6.51\times10^{-3}$, $9.217\times10^{-4}$, and $6.03\times10^{-5}$ on the validation grid. All QLG outputs were finite in these samples. The maximum of QLG-top is limited by \texttt{fp32} rounding in the setup, which differs between adjacent \texttt{fp32} values of $a$, so a denser sample in $a$ finds larger values than either the grid or the random samples. Evaluating every \texttt{fp32} value of $a$ in $[0.1,0.35]$ at $33$ abscissae in $0.002\le|v|\le0.4$ gives $6.98\times10^{-5}$ at $a=0.1728$, $|v|=0.020$, and Table~\ref{tab:deployed} quotes $7\times10^{-5}$ for this member. The reported maxima are empirical estimates from these checks, not certified global bounds.

\begin{figure}[t]
\centering
\includegraphics[scale=0.7]{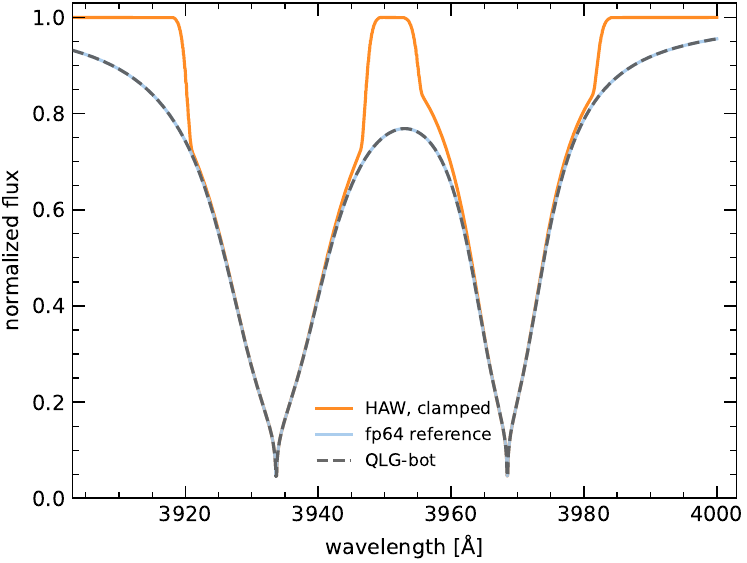}
\caption{Solar-dwarf normalized flux around Ca\,\textsc{ii} H\&K, calculated with HAW in \texttt{fp32}, the \texttt{fp64} reference kernel, and QLG-bot, whose maximum full-profile relative error over the validation domain is $6.5\times10^{-3}$ (Table~\ref{tab:deployed}). Negative HAW profile values were clamped to zero before their opacities entered the synthesis. QLG-bot, the least accurate of the three members, is visually indistinguishable from the reference here, whereas HAW suppresses the broad damping wings.}
\label{fig:cahk}
\end{figure}

\subsection{Timing and replay}

The M1 results used \texttt{gfortran 15}, \texttt{-O3 -mcpu=apple-m1}, without fast-math or whole-program optimization. Each kernel was called across a module boundary, as in the synthesis, and wrote one line window to an array. Timing includes the $a$-dependent setup but excludes opacity accumulation. Each reported value averages two independent passes, each the best of five sweeps over the recorded stream; the passes agree within $1.4\%$ on every row.

The x86-64 results used an AMD EPYC 9554 (Zen~4) node of the Viper cluster at the Max Planck Computing and Data Facility, with \texttt{gfortran 15.1.0 -O3 -march=native} and otherwise identical source, options, and procedure; the two passes agree within $0.8\%$. AVX-512 masking vectorises the branched \textsc{siu} and \textsc{moog} loops, evaluating the \texttt{glibc} exponential $\exp(-v^{2})$ even at far-wing points where its result is discarded and where the vectorised exponential is slow.

The instrumentation recorded $(n_w,a,\Delta v)$ for every line window. Wide windows were retained in full; narrow windows were sampled at strides of 257, 2053, and 4093 for the solar dwarf, red supergiant, and M dwarf. With $s_j$ the number of original windows represented by retained window $j$, the reported cost is
\begin{equation}
  \bar c_k =
  \frac{\sum_j s_j\,t_{kj}}
       {\sum_j s_j\,n_{w,j}},
  \label{eq:replay_cost}
\end{equation}
where $t_{kj}$ is the elapsed time for kernel $k$ on retained window $j$. The same ordered stream and weights were used for every kernel.

For $n_w\ge1000$, setup is amortized and every QLG member costs $1.6$--$1.9$\,ns per point on the M1. Narrow windows, with mean $n_w$ of $7.4$, $6.2$, and $3.2$ on the three atmospheres, distinguish the members. On the M dwarf, $68\%$ of narrow windows have $n_w=3$, so the scalar loop remainder dominates.

\subsection{Distribution of the synthesis flux residual}

The flux column of Table~\ref{tab:deployed} is set by a single pixel across the three syntheses. Table~\ref{tab:fluxdist} gives distributions for assessing whether the errors are tolerable in a particular analysis.